# Surface Topography: Metrology and Properties

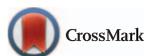

PAPER

OPEN ACCESS

# Model of rough surfaces with Gaussian processes

A Jawaid ⓘ and J Seewig ⓘ

Institute for Measurement and Sensor Technology, University of Kaiserslautern-Landau, Germany

E-mail: arsalan.jawaid@rptu.de

## Abstract

Surface roughness plays a critical role and has effects in, e.g. fluid dynamics or contact mechanics. For example, to evaluate fluid behavior at different roughness properties, real-world or numerical experiments are performed. Numerical simulations of rough surfaces can speed up these studies because they can help collect more relevant information. However, it is hard to simulate rough surfaces with deterministic or structured components in current methods. In this work, we present a novel approach to simulate rough surfaces with a Gaussian process (GP) and a noise model because GPs can model structured and periodic elements. GPs generalize traditional methods and are not restricted to stationarity so they can simulate a wider range of rough surfaces. In this paper, we summarize the theoretical similarities of GPs with auto-regressive moving-average processes and introduce a linear process view of GPs. We also show examples of ground and honed surfaces simulated by a predefined model. The proposed method can also be used to fit a model to measurement data of a rough surface. In particular, we demonstrate this to model turned profiles and surfaces that are inherently periodic.

## Abbreviations

The following abbreviations are used in this paper:

| | |
|---|---|
| ACF | Autocorrelation function |
| ACVF | Autocovariance function |
| AR | Auto-regressive |
| ARMA | Auto-regressive moving-average |
| FFT | Fast Fourier transformation |
| GP | Gaussian process |
| MA | Moving-average |
| PDF | Probability density function |
| PSD | Power spectral density |

## 1. Introduction

Functional requirements of technical workpieces (e.g. low flow resistance or hydrophobicity) are often traced back to geometric characteristics of the workpiece's surface [1, 2]. One approach to validate surface properties and their effects on these requirements is to measure both properties by real-world experiments, e.g. surface roughness by optical measurement instruments and flow resistance by wind tunnels. In contrast, model-based approaches can lead to significant speed-up and reduce costs in such investigations. For example, instead of collecting data from flow behaviors at different surface properties, computational fluid dynamic software can be used that takes surfaces with different properties as input and returns fluid dynamics simulations. To further automate these procedures, surfaces can also be simulated by a model-based approach. In this context, the simulation of rough surfaces has been mostly studied in research due to their importance mainly, but not only, in tribology.

To model rough surfaces that have multi-scale characteristics, fractal-based approaches may be suitable [3, 4]. Even though various rough surfaces have been modeled this way in the literature [5–7], the fractal properties are not always present [8, 9]. So, rough surfaces are typically modeled as stochastic processes (e.g. [10–12]). Nevertheless, fractal-based approaches are closely related to time series modeling [9].

Stochastic processes for rough surfaces assume that surfaces are described by two quantities. These are the probability density function (PDF) and the zero-mean autocovariance function (ACVF) [13]. Alternatives to the ACVF are the autocorrelation function (ACF)[1] and the power spectral density (PSD). We distinguish this modeling approach between two concepts. Some studies assume that the model quantities are given, and other studies model them from surface data.

In the former case, a predefined stationary[2] PSD or ACVF is mostly used to design a filter and filtering a white noise series results in a rough surface sample, e.g. [14]. The sampled surfaces have ACVFs or PSDs similar to the given ones. Since the filter design problem is generally a system of non-linear equations, the focus has been on solving the problem efficiently in the literature. For example, [15] applied the Newton method or [16] solved the system with the non-linear conjugate gradient method. [10] reformulated the non-linear equation system to a least squares problem and provided the gradient analytically. [17] proposed to approximately simulate stationary Gaussian processes (GPs) for rough surfaces. They also stated that stationary GPs are linear filters. Many other approaches [10, 11, 14, 18, 19] used the fast Fourier transform (FFT) algorithm due to its computational efficiency. To get surfaces with non-Gaussian PDFs, the white noise series can be transformed with a predefined skewness and kurtosis [11, 14, 20, 21].

One main challenge of these approaches is that the ACVF needs to be *a priori* known. For example, while these methods have allowed the simulation of random or ground surfaces, it is difficult to simulate complex rough surfaces because their ACVFs are often not known. Hence, there is a necessity to design an ACVF by surface data automatically. Therefore, there has been a lot of effort - not only in surface terminology - on ACVF or PSD design from data, e.g. [22, 23].

Modeling ACVFs from data is one of the reasons why auto-regressive moving-average (ARMA) processes [23] have been used extensively for rough surfaces [20, 24–26]. ARMA processes are linear shift-invariant filters and their filter coefficients have been designed with given ACVFs, PSDs, or surface data. [20] proposed to separate deterministic and stochastic components initially in surface data and to model the stochastic component with ARMA processes [24, 25]. Used methods to select the number of ARMA coefficients such that an ACVF can be automatically fitted to surface data. Even though it is effective to model the stochastic components of a surface with ARMA processes, they cannot model all stationary ACVFs. For example, a mere ARMA has led to problems for surfaces with deterministic components [25, 26]. And for surfaces with periodic components, a traditional ARMA approach is not suitable [23].

This paper uses GPs to model rough surfaces. With GPs it is not only possible to set an ACVF and simulate surfaces but also to select ACVFs using surface data automatically. Furthermore, GPs extend the space of possible surfaces by removing the restriction that rough surfaces are simulated by standard linear processes and even stationary stochastic processes. We introduce GPs in surface terminology, and we connect them with ARMA and linear processes. Finally, we simulate rough surfaces with predefined ACVFs and stationary ACVFs selected from the data.

## 2. Model of rough surfaces

### 2.1. Notations

Following [13], a rough surface is a real-valued random variable $Z(\boldsymbol{x}) \in \mathbb{R}$ with position $\boldsymbol{x} \in \mathcal{X} \subseteq \mathbb{R}^D$ and has zero-mean. A profile or a surface is considered, if $D = 1$ or $D = 2$, respectively. So, a rough surface with $\mathbb{E}[Z(\boldsymbol{x})] = 0$ is described by its PDF $p(z)$ and its ACVF

$$r(\boldsymbol{x}, \boldsymbol{x}') = \mathbb{E}[Z(\boldsymbol{x}) \cdot Z(\boldsymbol{x}')], \qquad (1)$$

or alternatively by its ACF

$$\rho(\boldsymbol{x}, \boldsymbol{x}') = \frac{\mathbb{E}[Z(\boldsymbol{x}) \cdot Z(\boldsymbol{x}')]}{(\mathbb{E}[Z(\boldsymbol{x}) \cdot Z(\boldsymbol{x})] \cdot \mathbb{E}[Z(\boldsymbol{x}') \cdot Z(\boldsymbol{x}')])^{\frac{1}{2}}}, \qquad (2)$$

If a further assumption is made that only stationary processes are considered, then the ACVF $r(\cdot)$ and the ACF $\rho(\cdot)$ are shift-invariant $\boldsymbol{\tau} = \boldsymbol{x}' - \boldsymbol{x}$ and the equations reduce to

$$r(\boldsymbol{\tau}) = \mathbb{E}[Z(\boldsymbol{x}) \cdot Z(\boldsymbol{x} + \boldsymbol{\tau})], \qquad (3)$$

$$\rho(\boldsymbol{\tau}) = \frac{\mathbb{E}[Z(\boldsymbol{x}) \cdot Z(\boldsymbol{x} + \boldsymbol{\tau})]}{\mathbb{E}[Z(\boldsymbol{x}) \cdot Z(\boldsymbol{x})]}. \qquad (4)$$

We consider only ACVFs because ACFs are normalized functions of their corresponding ACVFs. The importance of an ACVF is justified by the fact that it describes many properties, e.g. periodicity or random features of surfaces [9].

Regarding ACVF design, not every function is suitable to be a valid ACVF because it must be positive-semidefinite. In contrast, the PSD design has milder conditions [27]. To infer a stationary ACVF $r(\boldsymbol{\tau})$ from a PSD $S(\boldsymbol{f})$, the Wiener-Khintchine theorem [28, 29] is used

$$\begin{aligned} r(\boldsymbol{\tau}) &= \int_{\mathbb{R}^D} S(\boldsymbol{f}) \cdot e^{i 2\pi \boldsymbol{f}^\top \boldsymbol{\tau}} \, d\boldsymbol{\tau}, \\ S(\boldsymbol{f}) &= \int_{\mathbb{R}^D} r(\boldsymbol{\tau}) \cdot e^{-i 2\pi \boldsymbol{f}^\top \boldsymbol{\tau}} \, d\boldsymbol{f}. \end{aligned} \qquad (5)$$

Similarly, there exists a theorem for non-stationary ACVFs and generalized spectral densities, see [30]. The ACVF and the PSD are Fourier duals so a PSD can also be used to characterize rough surfaces.

---

[1] Following the notation common in time series analysis, ACFs are normalized ACVFs (see section 2.1).

[2] We denote weakly stationary as stationary.

## 2.2. Rough surfaces with Gaussian process and noise model

The proposed rough surface model can be split into two parts. The first part is a GP and the second part is a noise model. Thus, both quantities of rough surfaces (ACVF, PDF) are included in this approach. The ACVF is located in the GP and the PDF is in the noise model. To avoid incorrect notations, we use $G(\mathbf{x})$ to denote the latent output of the GP and $Z(\mathbf{x})$ to denote the output of the noise model or surface output. These two parts are described in the following.

Following [31] a GP is a stochastic process with Gaussian finite dimensional distribution. Therefore, a GP $G(\mathbf{x}) \in \mathbb{C}$ is fully characterized by a mean function $m(\mathbf{x})$ and an ACVF $k(\mathbf{x}, \mathbf{x}')$ denoted as

$$G(\mathbf{x}) \sim \mathcal{GP}(m(\mathbf{x}), k(\mathbf{x}, \mathbf{x}')),$$
$$m(\mathbf{x}) = \mathbb{E}[G(\mathbf{x})],$$
$$k(\mathbf{x}, \mathbf{x}') = \mathbb{E}[(G(\mathbf{x}) - m(\mathbf{x})) \cdot \overline{(G(\mathbf{x}') - m(\mathbf{x}'))}], \quad (6)$$

where $\overline{(\cdot)}$ denotes complex conjugation.

Setting the GP to have real values $G(\mathbf{x}) \in \mathbb{R}$ and its mean function to $m(\mathbf{x}) = 0$ is appropriate for rough surfaces (see section 2.1). Then the GP is described only by the ACVF for which $k(\mathbf{x}, \mathbf{x}') = r(\mathbf{x}, \mathbf{x}')$ holds (see (1)). The GP for rough surfaces is given as

$$G(\mathbf{x}) \sim \mathcal{GP}(0, r(\mathbf{x}, \mathbf{x}')). \quad (7)$$

In addition to the GP model, a noise model is specified by a PDF

$$Z(\mathbf{x}_n)|g(\mathbf{x}_n) \sim p(z(\mathbf{x}_n)|g(\mathbf{x}_n)), \quad (8)$$

e.g. a common noise model is the Gaussian noise $p(z(\mathbf{x}_n)|g(\mathbf{x}_n)) = \mathcal{N}(z(\mathbf{x}_n); g(\mathbf{x}_n), \sigma^2)$. The latent value $g(\mathbf{x}_n)$ is an observed GP at position $\mathbf{x}_n$. The final rough surface model is described as

$$G(\mathbf{x}) \sim \mathcal{GP}(0, r(\mathbf{x}, \mathbf{x}')),$$
$$Z(\mathbf{x}_n)|g(\mathbf{x}_n) \sim p(z(\mathbf{x}_n)|g(\mathbf{x}_n)). \quad (9)$$

*Simulation.* To simulate rough surfaces with a given model, the following two samplings are performed. Firstly, a finite set with $N$ positions $\mathcal{X}_N = \{\mathbf{x}_n\}_{n=1}^N$ must be defined. We choose, for example, a regularly meshed grid. With an $\mathcal{X}_N$, a GP $\mathbf{G} = \{G(\mathbf{x}), \mathbf{x} \in \mathcal{X}_N\}$ is sampled by a multivariate Gaussian distribution

$$\mathbf{G} \sim \mathcal{N}(\mathbf{G}; 0, \mathbf{R}), \quad (10)$$

where $[\mathbf{R}]_{ij} = r(\mathbf{x}_i, \mathbf{x}_j)$ is a $N \times N$ covariance matrix.

The latent surface $\mathbf{G}$ has the covariance matrix constructed from an ACVF. So, an exact sample from this multivariate Gaussian has the given ACVF. However, the sampling procedure is computationally expensive because the covariance matrix scales quadratically with the number of points to be simulated and therefore the Gaussian distribution usually becomes high-dimensional. For example, a surface with $512 \times 512$ points has a covariance matrix with a size of $262\,144 \times 262\,144$. Thus, this high-dimensional Gaussian distribution can be sampled, for example, with the reparameterization trick [32] followed by a matrix root decomposition. For the roughness simulations, we use LanczOs Variance Estimates [33] or Cholesky decomposition. Secondly, the sampling procedure to obtain a sampled surface $\mathbf{z} = \{z(\mathbf{x}), \mathbf{x} \in \mathcal{X}_N\}$ depends on the noise model. We used a Gaussian noise model that adds Gaussian white noise to the sampled GP $\mathbf{g}$. For other noise models, we refer to [32].

*Model selection.* When the ACVF or the PDF is only implicitly given, the model is not fully specified. This is true when the functions have some hyperparameters $\boldsymbol{\alpha}$ that are not set. For example,

$$r(\mathbf{x}, \mathbf{x}') = \sigma_k^2 \delta_{\mathbf{x},\mathbf{x}'},$$
$$p(z(\mathbf{x}_n)|g(\mathbf{x}_n)) = \mathcal{N}(z(\mathbf{x}_n); g(\mathbf{x}_n), \sigma^2), \quad (11)$$

with the hyperparameter set $\boldsymbol{\alpha} = \{\sigma_k, \sigma\}$ and $\delta_{\mathbf{x},\mathbf{x}'}$ the Kronecker delta.

These hyperparameters influence the simulated surface and are either chosen with *a priori* knowledge or using surface data. With surface data $\{\tilde{z}_k, \mathbf{x}_k\}_{k=1}^M$, the hyperparameters $\boldsymbol{\alpha}$ are usually optimized by maximizing the marginal likelihood $p(\{\tilde{z}_k\}_{k=1}^M|\{\mathbf{x}_k\}_{k=1}^M, \boldsymbol{\alpha})$ or by sampling with the marginal likelihood [31, 34]. After the hyperparameters are set, the model is determined and the simulation follows the aforementioned procedure.

In the previous paragraph, we assumed that the ACVF is implicitly given, e.g. (11). If a function for ACVF is not implicitly given, the model selection process using surface data selects a function within a positive-semidefinite function space. Methods to select an ACVF from data exploit the Fourier duality of PSD and ACVF in stationary [35, 36] or non-stationary settings [37, 38]. Although the function space is generalized without stationary assumptions, we consider methods that restrict the function space to stationary ACVFs only. Particularly, we apply the method proposed in [35] for turned profiles and surfaces (see section 3.3).

We use the framework GPyTorch [34] for model selection and simulation in this work.

## 2.3. Relation between Gaussian, ARMA and linear processes

ARMA processes have been widely used for model-based roughness. This section is intended to help better understand how GPs are theoretically related to ARMA processes. Furthermore, we introduce a continuous linear process representation of GPs here.

An ARMA$(p, q)$ process $\{G_x, x \in \mathbb{Z}\}$ is a discrete stochastic process defined by

$$G_x = \sum_{i=1}^{p} a_i G_{x-i} + \sum_{j=1}^{q} b_j \Delta B_{x-j} + \Delta B_x, \quad (12)$$

with a Brownian motion $\{B_x, x \in \mathbb{Z}\}$ or a white noise process $\{\Delta B_x, x \in \mathbb{Z}\}$. A moving-average (MA) and auto-regressive (AR) process are special cases of ARMA processes. An ARMA(0, q) process is a MA(q) process and an ARMA(p, 0) process is an AR(p) process. In contrast, GPs are continuous stochastic processes. After [31], a stationary ARMA(p,q) process has a stationary GP representation, whereby a discrete sampled process of this GP is ARMA(p,q). This is not necessarily true for AR(p) processes [39].

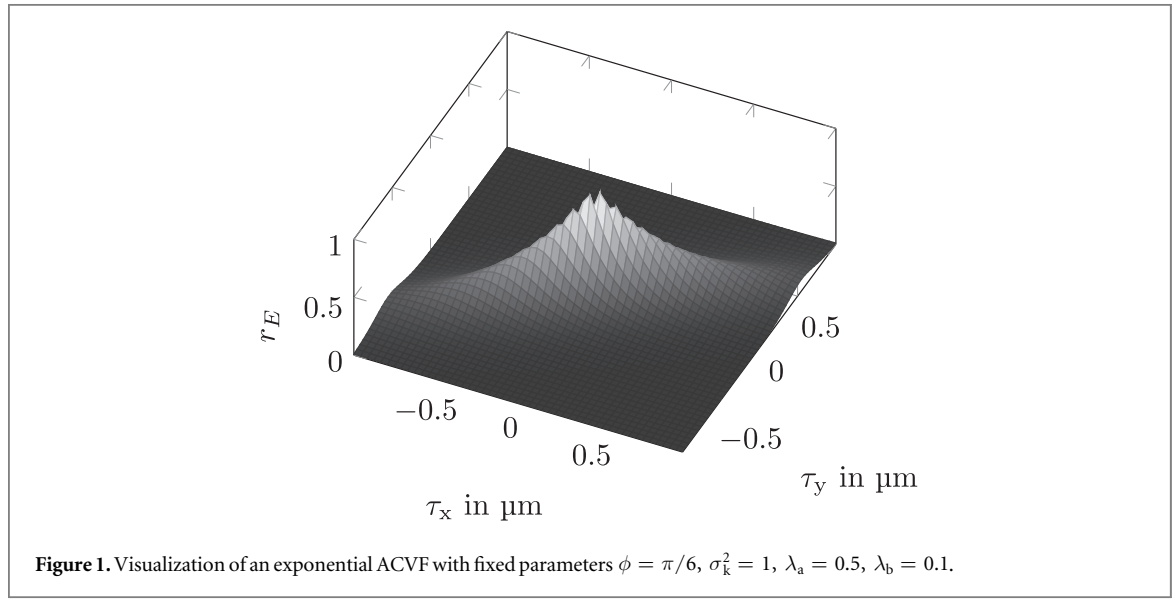

**Figure 1.** Visualization of an exponential ACVF with fixed parameters $\phi = \pi/6$, $\sigma_k^2 = 1$, $\lambda_a = 0.5$, $\lambda_b = 0.1$.

Moreover, both stochastic processes are connected through linear processes. So, a discrete stochastic process $\{G_x, x \in \mathbb{Z}\}$ of the form

$$G_x = \sum_{k=-\infty}^{\infty} \beta_k \Delta B_{x-k}, \quad (13)$$

is a linear process[3] [40]. From this equation, it is clear that MA(q) processes are casual linear processes with finite coefficients. It is also true that a stationary AR(p) process or a stationary ARMA(p, q) process has a casual linear process representation [23]

$$G_x = \sum_{k=0}^{\infty} \tilde{\beta}_k \Delta B_{x-k}. \quad (14)$$

In [41–43] it is stated that GPs can be represented as a convolution between a feature map $\beta: \mathbb{R}^D \mapsto \mathbb{C}$ and a white noise process, and therefore, GPs are continuous MA processes. More generally, GPs can be represented as a stochastic differential equation defined as

$$dG(\boldsymbol{x}) = \beta(\boldsymbol{\kappa})\, dB(\boldsymbol{x} - \boldsymbol{\kappa}) + dm(\boldsymbol{x}), \quad (15)$$

where $B(\boldsymbol{x})$ is a multivariate Brownian motion. So, a GP

$$G(\boldsymbol{x}) - m(\boldsymbol{x}) = \int_{\mathbb{R}^D} \beta(\boldsymbol{\kappa})\, dB(\boldsymbol{x} - \boldsymbol{\kappa}), \quad (16)$$

with ACVF

$$k(\boldsymbol{x}, \boldsymbol{x}') = \int_{\mathbb{R}^D} \beta(\boldsymbol{x} - \boldsymbol{k}) \overline{\beta(\boldsymbol{x}' - \boldsymbol{k})}\, d\kappa, \quad (17)$$

can be viewed not only as a non-casual and continuous MA process but also as a continuous linear process.

---
[3] Formally, it is a generalized linear process.

Consequently, ARMA processes are special cases in GPs.

## 3. Results

We demonstrate the GP approach for different surfaces in this part, whereby we assumed Gaussian noise models with a constant parameter ($\sigma^2 = 1$) in all cases. We simulated ground surfaces and honed surfaces by manually designing ACVFs in the first two subsections. Also, we used the method in [35] to automatically reconstruct ACVFs of turned surfaces from measurement data in the last subsection.

### 3.1. Ground surfaces

Ground surfaces have been simulated in various publications [44–46]. The texture of these surfaces can have an arbitrary orientation that is created by a grinding process, which can be performed in an arbitrary direction. For ground surfaces, an exponential ACVF has been used which is stationary and rotates its inputs by $\phi$

$$\begin{aligned} r_E(\boldsymbol{\tau}; \phi) &= \sigma_k^2 \exp\left(-(\boldsymbol{\tau}'^{\top} \boldsymbol{\Lambda}^{-2} \boldsymbol{\tau}')^{\frac{1}{2}}\right), \\ \boldsymbol{\tau}' &= \boldsymbol{T}_\phi \boldsymbol{\tau}^{\top}, \end{aligned} \quad (18)$$

where $\boldsymbol{T}_\phi$ is the inverse rotation matrix and $\boldsymbol{\Lambda}$ is a diagonal matrix with lengthscales $\lambda_a$, $\lambda_b$ as diagonal elements. In order for the ground surface to have an orientation, the lengthscales must be unequal. Also, it should be noted that a linear transformation with a matrix does not make the ACVF invalid.

In figure 1, an exponential ACVF is visualized with $\phi = \pi/6$. This generates high correlation values along the rotated $x$-axis and therefore leads to grinding grooves parallel to this axis. A ground surface with 2000 × 2000 points has been simulated according to section 2.2 (see figure 2) with similar parameters as the ACVF in figure 1. Because of the high-dimensional Gaussian distribution, we sampled the latent surface with an approximation [33]. To indicate an approximation error, we sampled 50 latent ground surfaces each with $100 \times 100$ points and hyperparameters according to figure 2. Then, we computed the mean absolute error between the given covariance matrix and the unbiased sample covariance matrix element-wise. The error was $2.763 \cdot 10^{-1}$. Note that the error in exact sampling with Cholesky decomposition was $2.536 \cdot 10^{-1}$, so both errors have a comparable magnitude.

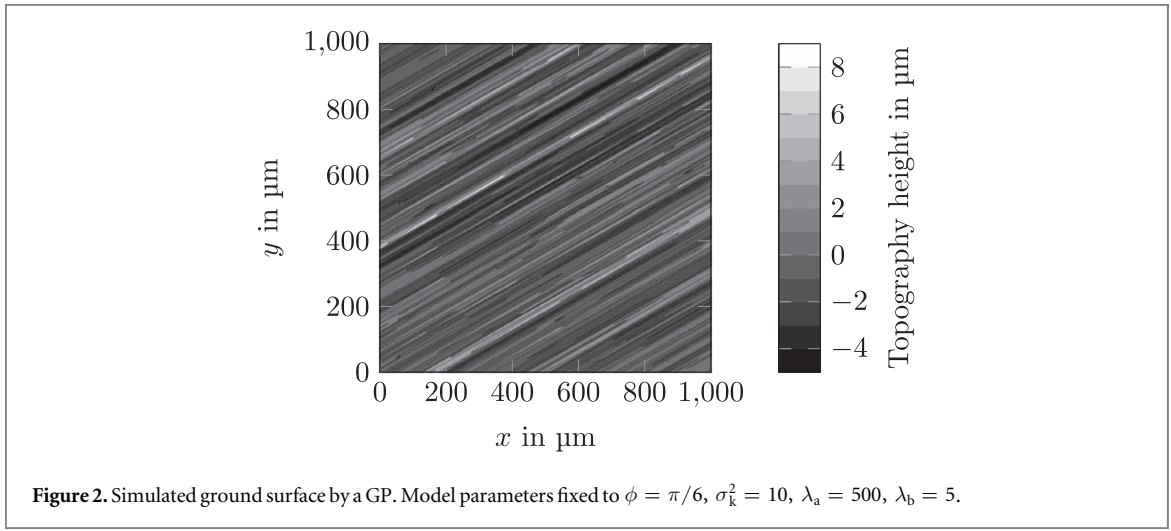

**Figure 2.** Simulated ground surface by a GP. Model parameters fixed to $\phi = \pi/6$, $\sigma_k^2 = 10$, $\lambda_a = 500$, $\lambda_b = 5$.

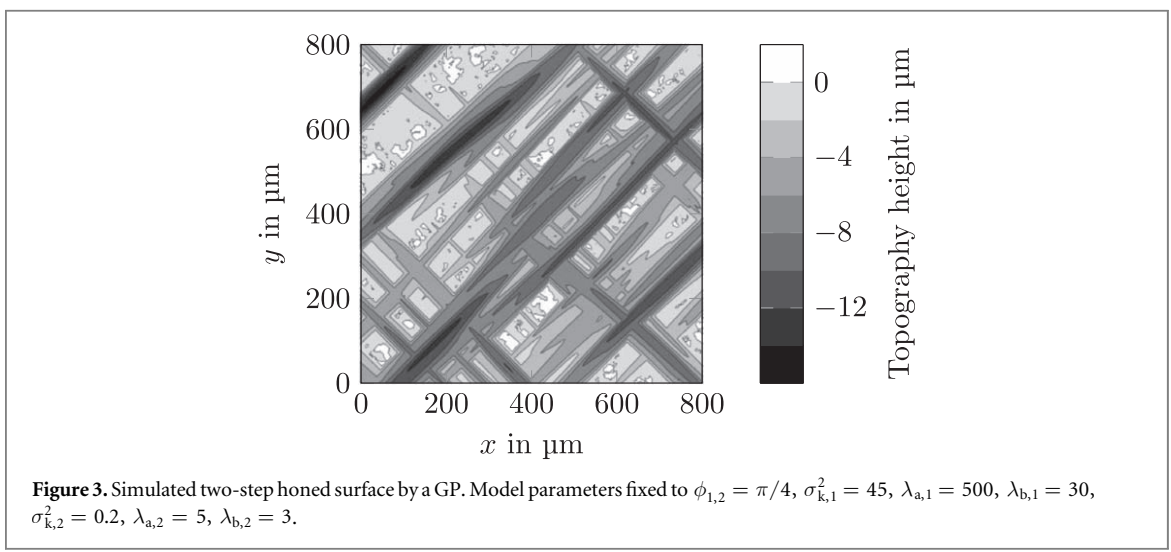

**Figure 3.** Simulated two-step honed surface by a GP. Model parameters fixed to $\phi_{1,2} = \pi/4$, $\sigma_{k,1}^2 = 45$, $\lambda_{a,1} = 500$, $\lambda_{b,1} = 30$, $\sigma_{k,2}^2 = 0.2$, $\lambda_{a,2} = 5$, $\lambda_{b,2} = 3$.

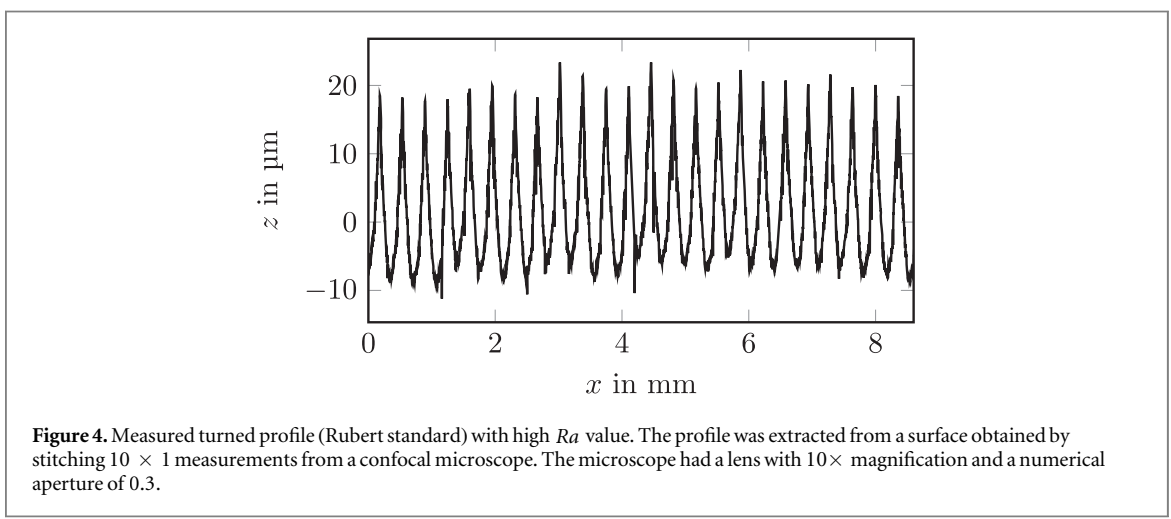

**Figure 4.** Measured turned profile (Rubert standard) with high *Ra* value. The profile was extracted from a surface obtained by stitching $10 \times 1$ measurements from a confocal microscope. The microscope had a lens with $10\times$ magnification and a numerical aperture of 0.3.

### 3.2. Honed surfaces

Grinding tools are used in honing processes. In contrast to grinding, torques of honing tools act parallel to feed motions. The tool movements periodically change the direction in a honing process. Hence, a honed surface can be assumed as a superposition of ground surfaces as discussed in [45, 46].

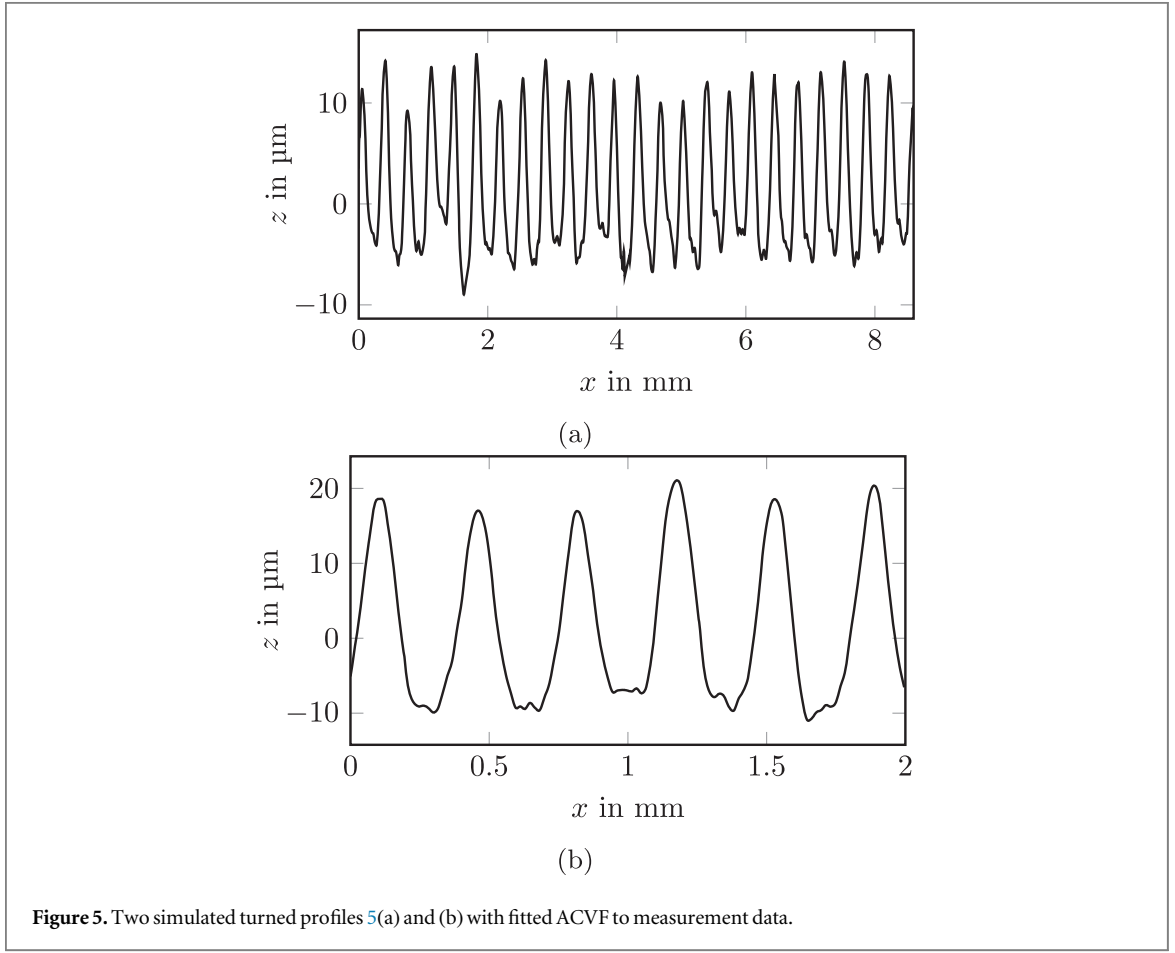

**Figure 5.** Two simulated turned profiles 5(a) and (b) with fitted ACVF to measurement data.

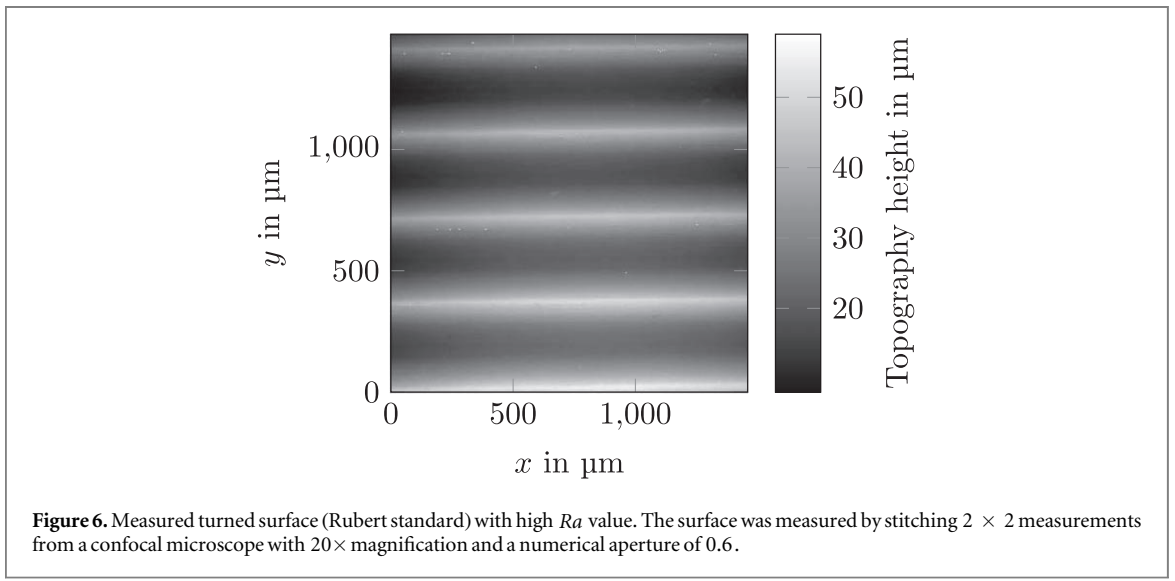

**Figure 6.** Measured turned surface (Rubert standard) with high *Ra* value. The surface was measured by stitching 2 × 2 measurements from a confocal microscope with 20× magnification and a numerical aperture of 0.6.

Thus, an exponential ACVF (18) is used for a rough surface created by a one-step honing process. In this simulation, two ground surfaces $\{z_\phi, z_{-\phi}\}$ with mirrored grinding groves are realized and then a simulated one-step honed surface is obtained according to

$$z_h = \{\min(z_\phi(\boldsymbol{x}), z_{-\phi}(\boldsymbol{x})), \boldsymbol{x} \in \mathcal{X}_N\}. \quad (19)$$

A *P*-step honed surface is achieved with the help of *P* simulations of one-step honed surfaces

$$z_h = \{\min(z_{h,1}(\boldsymbol{x}),...,z_{h,P}(\boldsymbol{x})), \boldsymbol{x} \in \mathcal{X}_N\}. \quad (20)$$

The parameter sets of these individual simulations $\{z_{h,p}\}_{p=1}^{P}$ are independent of each other because single process steps in the multi-step honing process are also independent. A two-step honed surface with size

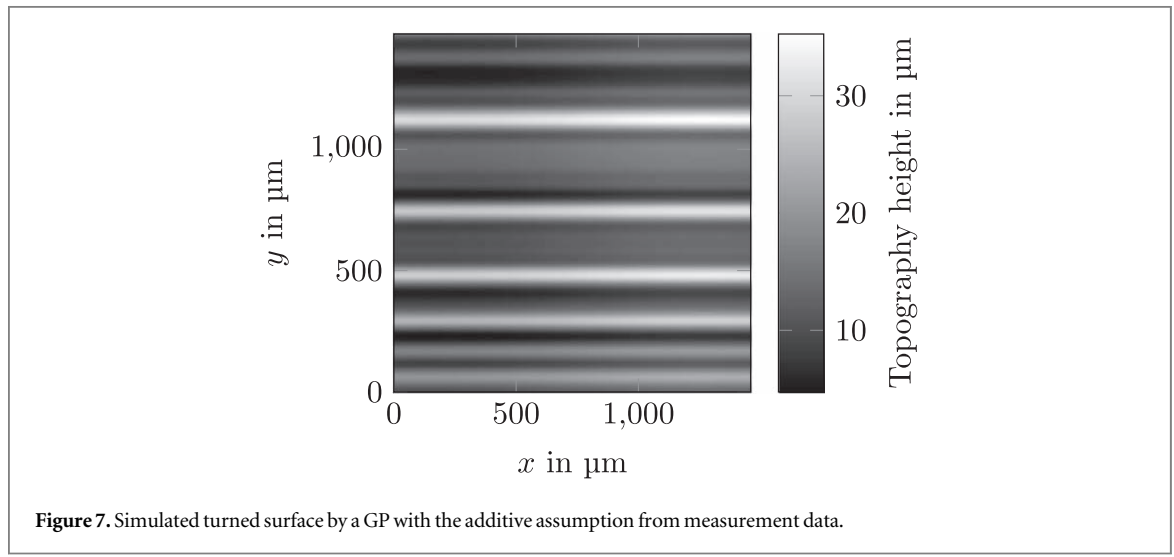

**Figure 7.** Simulated turned surface by a GP with the additive assumption from measurement data.

512 × 512 has been simulated with this procedure in figure 3. Due to the mirrored $\phi$ in underlying ground surfaces, the characteristic cross-structure of a honed surface is obtained.

### 3.3. Turned surfaces

To simulate honed surfaces, *a priori* knowledge has been used. However, if there is no *a priori* knowledge or if it is difficult to infer *a priori* knowledge, it is challenging to select an ACVF that reproduces the surface properties. In this subsection, we show how the proposed GP approach can model an ACVF of a turned surface from data since standard ARMA processes are not capable to infer ACVFs of such surfaces [23]. Although we use the proposed method in [35], which searches for an appropriate ACVF only in stationary space, there exist approaches to search also in non-stationary spaces, e.g. [37].

Initially, we consider only the simulation of turned profiles by measurement data that are shown in figure 4. [35] leveraged the relation in (5) to shift the problem to model PSDs instead of ACVFs. In this method, the PSD is modeled as a symmetrized and weighted sum of Gaussians which inverse Fourier transformation is the spectral mixture ACVF given as

$$r_{\text{SM}}(\boldsymbol{\tau}) = \sum_{q=1}^{Q} w_q \cos(2\pi \boldsymbol{\tau}^\top \boldsymbol{\mu}_q) \exp\left(-2\pi^2 \boldsymbol{\tau}^\top \boldsymbol{\Sigma}_q \boldsymbol{\tau}\right),$$
(21)

with $w_q$ the weight, $\boldsymbol{\mu}_q$ the mean and $\boldsymbol{\Sigma}_q$ the diagonal covariance matrix of the $q$-th Gaussian.

This approach can model all stationary ACVFs but it requires a good initialization of the hyperparameters [35, 36]. To initialize these hyperparameters, we first estimate PSDs from data by the Welch method [22] and by a periodogram. Secondly, Gaussian mixtures are fitted with inverse transform sampling into these noisy estimates multiple times, whereby these sets of fitted parameters are candidates for the initial values. To finally decide which set of initial values to pick, the marginal likelihood is used as a criterion. After this initialization procedure, the marginal likelihood is optimized with respect to the hyperparameters.

After the PSD has been modeled with $Q = 5$, two simulated turned profiles are visualized in figure 5. In these simulations, not only the periodicity has been extracted from the measurement data of a turned profile by modeling a stationary ACVF automatically but also the characteristics of wide dales and narrow hills have been modeled.

To obtain turned surfaces instead of turned profiles, this approach can be used, e.g. in its multivariate variant or with the additivity assumption. Turned surfaces have their main features in one dimension. In the other dimension, these features do not change significantly. Across this dimension, it shows rather a constant behavior, which can be described with, e.g. an exponential ACVF with large values for its lengthscale. For this reason, we modeled the total ACVF of a turned surface with an additivity assumption. The total ACVF consists of two separate ACVFs assigned to each dimension

$$r(\boldsymbol{\tau}) = r_x(\tau_x) + r_y(\tau_y). \quad (22)$$

We assumed only the additive behavior and modeled both ACVFs each with (21) and $Q = 10$ to show an automatic ACVF selection.

To model these two ACVFs, we have used measurement data (see figure 6). Similar to profile simulations, the periodicity and other features of the turned surface, like wide dales and narrow hills, are modeled in the ACVF and reproduced in the simulation (see figure 7). However, this approach falsely simulates surfaces with additional short wavelengths, which is specifically observed in the dales. The superposition with these wavelength components leads to a decreasing periodicity in the simulation. The cause of the errors is due to the modeling approach of the ACVF. Modeling the ACVF approach with Gaussian mixtures (21) results in an optimization problem that is likely to be multi-modal [35]. Other models (e.g. hierarchical

GPs [36] or non-stationary ACVF models [47]) might be more stable and appropriate here. So, the errors are mainly due to the choice of modeling and the resulting local optima despite our initialization of the parameters.

## 4. Conclusion

This work introduced a GP and noise model approach to model rough surfaces. We applied this approach to simulate ground and honed surfaces. In addition, we showed that GPs can model rough surfaces with strong periodic components automatically from measurement data. Particularly, we applied spectral mixture ACVFs [35] to model turned profiles and surfaces. The model produced better simulations for profiles than for surfaces due to the choice of the ACVF model. Nevertheless, the results show that the proposed approach can simulate rough surfaces with long correlation lengths that are not possible with ARMA processes or other digital filter methods. Moreover, if latent surfaces can be sampled exactly from the multi-variate Gaussian distribution, their ACVF is consistent with the predefined ACVF.

With regard to the high-dimensional Gaussian distribution, this approach has its limitations. Sampling from this distribution is memory and computationally expensive compared to conventional simulation methods for roughness. Fitting the model to data is also memory and computationally intensive. Additionally, stationary GPs with Gaussian noise models cannot always appropriately describe the rough surface. This is specifically if the machined roughness exhibits strong non-Gaussian distribution. However, we do not restrict conceptually our approach to stationary GPs and Gaussian noise models.

Future work can go in several directions. One research area is to leverage GP's linear process view, e.g. to sample from a GP more efficiently [18], sample non-Gaussian surfaces with non-Gaussian white noise [21] or simulate rough surfaces with predefined roughness parameters [16]. Another research field is the modeling and simulation of non-stationary surfaces from data [37, 47]. Optical measurements may contain non-measured or spurious points due to the surface's optical suitability (e.g. high gradients), so an additional application of this model is to interpolate these points in measurement data [48].

## Acknowledgments

We thank Viktor Follmann and Samuel Schmidt for their helpful discussions. This research was funded by the Deutsche Forschungsgemeinschaft (DFG, German Research Foundation) - 252408385 - IRTG 2057.

## Data availability statement

The data that support the findings of this study are available upon reasonable request from the authors.

## ORCID iDs

A Jawaid  https://orcid.org/0000-0001-9951-4164
J Seewig  https://orcid.org/0000-0002-1420-1597